\documentclass[preprint,11pt]{elsarticle}
\usepackage{amsmath}
\usepackage{amsfonts}   % if you want the fonts
\usepackage{amssymb}    % if you want extra symbols
\usepackage{amsthm}
\usepackage{subfigure}

\makeatletter
\def\ps@headings{%
\def\@oddhead{\mbox{}\scriptsize\rightmark \hfil \thepage}%
\def\@evenhead{\scriptsize\thepage \hfil \leftmark\mbox{}}%
\def\@oddfoot{}%
\def\@evenfoot{}}
\makeatother
\journal{Discrete Applied Mathematics}

\begin{document}

\begin{frontmatter}

\newtheorem{thm}{Theorem}[section]
\newtheorem{cor}[thm]{Corollary}
\newtheorem{lem}[thm]{Lemma}
\newtheorem{prop}[thm]{Proposition}

\newtheorem{rem}[thm]{Remark}

\newtheorem{ex}[thm]{Example}

\newtheorem{defi}[thm]{Definition}

\title{On Vertex Attack Tolerance in Regular Graphs}


\author{Gunes Ercal}

\address{Southern Illinois University Edwardsville}

\begin{abstract}
We have previously introduced vertex attack tolerance (VAT), denoted mathematically as $\tau(G) =  \min_{S \subset V} \frac{|S|}{|V-S-C_{max}(V-S)|+1}$ where  $C_{max}(V-S)$ is the largest connected component in $V-S$, as an appropriate mathematical measure of resilience in the face of targeted node attacks for arbitrary degree networks.  A part of the motivation for VAT was the observation that, whereas conductance, $\Phi(G)$, captures both edge based and node based resilience for regular graphs, conductance fails to capture node based resilience for heterogeneous degree distributions.  We had previously demonstrated an upper bound on VAT via conductance for the case of $d$-regular graphs $G$ as follows: $\tau(G) \le d\Phi(G)$  if $\Phi(G) \le \frac{1}{d^2}$ and  $\tau(G) \le d^2\Phi(G)$ otherwise.  In this work, we provide a new matching lower bound: $\tau(G) \ge \frac{1}{d}\Phi(G)$.  The lower and upper bound combined show that $\tau(G) = \Theta(\Phi(G))$ for regular constant degree $d$ and yield spectral bounds as corollaries.
\end{abstract}

\end{frontmatter}

\section{Introduction}
Conductance\cite{ChungSpectraBook} is a well studied and important measure of network resilience in the face of edge failures or attacks.  Combinatorial conductance or edge based conductance\cite{SinclairJerrum,ChungSpectraBook} is defined as
\begin{eqnarray*}
\Phi(G) &= \min_{S \subset V, Vol(S) \le Vol(V)/2} \{ \frac{|Cut(S,V-S)|}{Vol(S)} \} \\
 &= \min_{S \subset V,  Vol(S) \le Vol(V)/2} \{ \frac{|Cut(S,V-S)|}{\delta_S|S|} \} 
\end{eqnarray*}
where $|Cut(S,V-S)|$ is the size of the cut separating $S$ from $V-S$, $Vol(S)$ is the sum of the degrees of vertices in $S$, and $\delta_S$ is the \emph{average} degree of vertices in $S$.  A reason for the importance of conductance is in the intimate relationship it shares with spectral gap and mixing time via \emph{Cheeger's inequality} \cite{ChungSpectraBook,SinclairJerrum}:
\begin{thm}\label{thm:cheeger}
Given a connected $d$-regular graph $G = (V,E)$, let $\lambda_2$ denote the second largest eigenvalue of $G$'s normalized adjacency matrix.  Then,
\begin{equation}\label{eqn:cheeger}
\frac{\Phi(G)^2}{2} \le 1 - \lambda_2 \le 2\Phi(G)
\end{equation}
\end{thm}
Given the general inapproximability of conductance to any constant degree, Cheeger's inequality nonetheless yields good asymptotic bounds on the conductance of certain infinite graph families via spectral gap.  An important characterization in this regard has been that of \emph{expander families} which are $d$-regular families of graphs that maintain excellent resilience properties despite keeping constant degree $d$ \cite{Alon86}.  And, although conductance implicitly refers to edge based resilience, it is known to accurately represent node-based resilience for regular graphs as well.

However, conductance neither captures nor approximates node-based resilience in heterogeneous degree graphs, the most extremal example of which is the star family of graphs.  Star graphs exhibit maximal conductance while being minimally resilient to node attacks, particularly that targeting the single central node and resulting in solely isolated nodes.  With the need for a measure that behaves similarly to conductance in the case of regular graphs while also appropriately capturing node based resilience for heterogeneous degree graphs, the notion of \emph{vertex attack tolerance} (VAT, denoted by $\tau(G)$) was introduced by the authors in \cite{CASResSF,MattaBE14} and defined as:
\begin{equation*}
\tau(G) = \min_{S \subset V, S \neq \emptyset} \{\frac{|S|}{|V-S-C_{max}(V-S)|+1 } \} 
\end{equation*}
where $C_{max}(V-S)$ is the largest connected component in $V-S$.  In addition to comprehensive comparisons with other resilience measures for arbitrary degree graphs, the following upper bound was presented for VAT in terms of conductance in the case of $d$-regular graphs:\cite{MattaBE14}
\begin{thm}\label{thm:vatcond}
For any non-trivial $d$-regular graph $G = (V,E)$, if $\Phi(G) \le \frac{1}{d^2}$ then $\tau(G) < d\Phi(G)$.
\end{thm}
Note that due to both conductance and VAT being normalized measures in $(0,1]$, this can be unconditionally phrased as $\tau(G) < d^2\Phi(G)$ for $d$-regular non-trivial graphs $G$.

The primary contribution of the present work is to present a matching lower bound for VAT in terms of conductance for $d$-regular graphs when the conductance is not too high.  Namely, we present and prove the following theorem:
\begin{thm}\label{thm:condvat}
For any non-trivial $d$-regular graph $G = (V,E)$, $\Phi(G) < d\tau(G)$.
\end{thm}

Applying Cheeger's inequality to both the previous bound and our new bound results in the following corollary:
\begin{cor}\label{cor:cheegervat}
Given a connected $d$-regular graph $G = (V,E)$, let $\lambda_2$ denote the second largest eigenvalue of $G$'s normalized adjacency matrix.  Then,
\begin{equation}
\frac{\tau(G)^2}{2d^4} \le 1 - \lambda_2 \le 2d\tau(G)
\end{equation}
Furthermore, if $\Phi(G) \le \frac{1}{d^2}$, then 
\begin{equation}
\frac{\tau(G)^2}{2d^2} \le 1 - \lambda_2 \le 2d\tau(G)
\end{equation}
\end{cor}

A secondary contribution of this work is to generalize VAT in two ways, the first via parametrization and the second via generalization to node-weighted graphs.  We first present this secondary contribution and then prove our primary contribution.

\section{Definitions and Preliminaries}
Throughout this work we assume that the graph $G = (V,E)$ whose resilience is under consideration is connected and undirected.  The resilience of any disconnected graph is otherwise zero. \footnote{To non-trivially define the resilience of a disconnected network one may restrict consideration to the largest connected component WLOG.}  And, for any directed graph we may instead consider the underlying undirected graph without edge directionality.  Therefore, in the context of this work, we refer to connected, undirected graphs $G = (V,E)$ with more than one node ($|V| \ge 2$) as \emph{non-trivial}.

Set-vertex attack tolerance is denoted as \cite{MattaBE14}
\begin{equation*}
\tau_S(G) =\frac{|S|}{|V-S-C_{max}(V-S)|+1 }
\end{equation*}
so that clearly $\tau(G) = \min_{S \subset V} \tau_S(G)$ and correspondingly
\begin{equation*}
S(\tau(G)) = argmin_{S \subset V} \tau_S(G)
\end{equation*}
Similary for set-conductance:
\begin{equation*}
\Phi_S(G) = \frac{|Cut(S,V-S)|}{\delta_S|S|}
\end{equation*}
so that clearly $\Phi(G) = \min_{S \subset V, |Vol(S)| \le |Vol(V)|/2} \Phi_S(G)$ and correspondingly
\begin{equation*}
S(\Phi(G)) = argmin_{S \subset V, |Vol(S)| \le |Vol(V)|/2} \tau_S(G)
\end{equation*}

Let us denote the subgraph of a graph $G = (V,E)$ that is induced by a vertex set $S \subset V$ as $G_S = (S, E_S)$.

The \emph{normalized adjacency matrix} of a graph $G = (V,E)$ is the $|V|$ by $|V|$ matrix $A$ where $A_{u,v} = 0$ if $\{ u,v \} \notin E$ and $A_{u,v} = \frac{1}{d_u}$ where $d_u$ is the degree of $u$ otherwise if $\{ u,v \} \in E$.  Note that the normalized adjacency matrix of a graph is identical to the probability transition matrix, or Markov chain, of the natural random walk.

\subsection{A Useful Generalization}
A particularly useful generalization of vertex attack tolerance that we introduce here which allows for linear weighting parameters $\alpha, \beta$ is the following, which we call $(\alpha,\beta)$-vertex attack tolerance:
\begin{equation*}
\tau_{\alpha,\beta}(G) = \min_{S \subset V, S \neq \emptyset} \{\frac{\alpha|S|+\beta}{|V-S-C_{max}(V-S)|+1 } \} 
\end{equation*}
Clearly VAT is identical to $(1,0)$-VAT.  We will later see that $\tau_{1,1}$, which nominally appears quite similar to VAT, will also be helpful.

We also wish to introduce one further generalization of vertex attack tolerance which allow us to discuss a type of vertex-weighted graph $G$ which has costs $c(x)$ and values $v(x)$ associated with each vertex $x \in V$.   Specifically, the most general graph context concerned here is that of undirected, connected graphs $G = (V, E, c, v)$ such that $c, v$ are positive real-valued functions on the vertex set $V$.  If $c$ is specified without $v$ being specified, then we will assume that $c = v$, and similarly for the case that $v$ is specified without $c$ being specified.  If neither $c$ nor $v$ are specified, then the assumption is that $c(x) = v(x) = 1$ for all $x \in V$.  Having clarified the context of such cost-value node-weighted graphs $G$, the following is the appropriate generalization of VAT:
\begin{equation*}
\tau(G) = \min_{S \subset V, S \neq \emptyset} \{\frac{\sum_{x \in S} c(x)}{1+ \sum_{y \in V} v(y) - \sum_{y \in S+C_{max}(V-S)} v(y)} \} 
\end{equation*}
Note that this generalization is consistent with the original definition of VAT in the originally considered case of unweighted graphs $G$.  Moreover, applying other variations and generalizations to cost-value weighted graphs is straightforward.  As an example, we may consider $(\alpha,\beta)$-VAT of cost-value node weighted graphs:
\begin{equation*}
\tau(G) = \min_{S \subset V, S \neq \emptyset} \{\frac{\alpha(\sum_{x \in S} c(x))+\beta}{1+ \sum_{y \in V} v(y) - \sum_{y \in S+C_{max}(V-S)} v(y)} \} 
\end{equation*}

\subsection{Mathematical Preliminaries}

Let us note the following preliminary observation:
\begin{rem}\label{rem:vatbound}
For nontrivial $G = (V,E)$, $0 < \tau(G) \le 1$, and $C_{max}(V-S(\tau(G))) \neq \emptyset$.
\end{rem}
The first bound follows from the non-emptiness of $S$ by definition of $\tau$, in addition to the fact that, for any vertex $v \in V$, $\tau(G) \le \tau_{\{ v \} }(G) = \frac{1}{|V - \{ v \} - C_{max}|+1} \le 1$.  The non-emptiness of the largest remaining connected component follows from the fact that the only way that $C_{max}$ can be empty is by taking $S = V$, but such $S$ cannot achieve as low a set-VAT as that achieved by a single node, and therefore cannot be the set corresponding to VAT.

The following similar bound is well-known for conductance:
\begin{rem}\label{rem:condbound}
For nontrivial $G = (V,E)$, $0 < \Phi(G) \le 1$.
\end{rem}

For the proofs to come, the following inequality will prove useful:
\begin{equation}\label{eqn:fracmid}
\forall a,b,x,y > 0, \; \frac{a}{x} < \frac{b}{y} \; \rightarrow \; \frac{a}{x} < \frac{a+b}{x+y} < \frac{b}{y}
\end{equation}
Even more useful is a corollary of this inequality that follows by induction:
\begin{cor}\label{cor:fracseries}
Let $n > 0$ be a natural number, and for each natural number $i$ from $1$ to $n$, let positive numbers $a_i$ and $b_i$ be given.  Moreover, let $c$ be any real number that satisfies $c \le \min_{1 \le i \le n} \frac{a_i}{b_i}$.  Then, the following is true:
\begin{equation}
c \le \frac{\sum_{1 \le i \le n} a_i}{\sum_{1 \le i \le n} b_i}
\end{equation}
\end{cor}

Additionally, we note the following lemma regarding conductance in $d$-regular graphs\cite{MattaBE14}:
\begin{lem}\label{lem:sconn}
Given a connected, undirected $d$-regular graph $G = (V,E)$, there exists a set $S$ such that $S = S(\Phi(G))$ and the induced subgraph $G_S$ is connected.
\end{lem}

\section{Lower-Bounding VAT}

\begin{proof}[\textbf{Proof of Theorem \ref{thm:condvat}}]
Let $S = S(\tau(G))$, which is non-empty by definition. Furthermore, let $T = C_{max}(V-S)$, which is also non-empty by non-triviality of $G$ and Remark \ref{rem:vatbound}.  Consider the set conductance of $T$, namely $\Phi_T(G)$.  There are two situations which we must consider separately:
\begin{enumerate}
\item $|T| > \frac{|V|}{2}$
\item $|T| \le \frac{|V|}{2}$
\end{enumerate}
Let us first consider the first situation.  In that case, $\Phi_T(G) = \frac{|Cut(T,V-T)|}{d|V-T|}$.  However, because all edges of $Cut(T,V-T)$ must be adjacent to $S$, we know that $|Cut(T,V-T)| \le d|S|$.  Moreover, clearly, $|V-T| \ge |V-S-T+1|$.  Combining these facts, we obtain that:
\begin{equation}
\Phi(G) \le \Phi_T(G) \le \frac{d|S|}{|V-S-T+1|} = d\tau(G)
\end{equation}
This finishes the proof for the case that $T$ is the majority of the nodes.  Therefore, let us finally consider the case that $T$ does not comprise the majority of the vertices:

Let us denote $q$ as the number of connected components of the induced subgraph $G_{V-S-T}$, and, furthermore, denote each such remaining component by $C_i, \forall 1 \le i \le q$.  Because $T$ itself is the \emph{largest} connected component of induced subgraph $G_{V-S}$ by definition, and $T < \frac{|V|}{2}$ by assumption, we know further that each $|C_i| <  \frac{|V|}{2}$.  It will be convenient to denote $C_{q+1} = T$.  Therefore, 
\begin{equation}\label{eqn:ccphi}
\forall 1 \le i \le q+1, \Phi_{C_i}(G) = \frac{|Cut(C_i,V-C_i)|}{d|C_i|}
\end{equation}
Now, note the following facts due to the action of $S$ and the definition of connected components:
\begin{enumerate}
\item $d|S| \ge |Cut(T,V-T)| + \sum_{i=1}^q |Cut(C_i,V-C_i)| =  \sum_{i=1}^{q+1} |Cut(C_i,V-C_i)|$ 
\item $|V-S-T+1| \le |V-S| = |T| + \sum_{i=1}^q |C_i| =  \sum_{i=1}^{q+1} |C_i|$
\end{enumerate}
The second fact is clear from the definition of connected components as a partition.  The first fact follows from the action of $S$: All edges crossing the boundary of a connected component must be adjacent to $S$.  Combining the two facts, we obtain that
\begin{equation}
\tau(G) = d\frac{|S|}{d|V-S-T+1|} \ge  \frac{\sum_{i=1}^{q+1} |Cut(C_i,V-C_i)|}{\sum_{i=1}^{q+1} d|C_i|}
\end{equation}
Now, note that each term of the sum in the numerator is a numerator of $\Phi_{C_i}(G)$ whereas each term of the sum in the denominator is a corresponding denominator of $\Phi_{C_i}(G)$.  Applying Corollary \ref{cor:fracseries} with $c = \Phi(G)$, with $a_i = |Cut(C_i,V-C_i)|$, and with $b_i =  d|C_i|$, we obtain that $\tau(G) \ge \Phi(G)$, completing the proof.
\end{proof}

\bibliographystyle{plain}
\bibliography{spectralVAT}

\end{document}